\documentclass[aps,pra,twocolumn,preprintnumbers,notitlepage,tightenlines,preprintnumbers,amsmath,amssymb]{revtex4-2}

\usepackage{subfigure}
\usepackage{graphicx,amsmath}
\usepackage{epsf}
\usepackage{natbib}
\usepackage{xcolor}

\usepackage{graphicx}
\usepackage{bm}
\usepackage{dsfont}

\begin{document}

\title{Memory Device for Photons by exploiting Brillouin Interactions in Nanowires}
\author{Hashem~Zoubi}
\email{hashemz@hit.ac.il}
\affiliation{Department of Physics, Faculty of Sciences, Holon Institute of Technology, Holon 5810201, Israel}
\date{19 December 2025}

\begin{abstract}
Memory devices for single photons are notable components for quantum information processing and quantum communications. The present study investigates the possibility of achieving storage of light at the level of single photons inside nanofibers by exploiting stimulated Brillouin scattering. We present first the standard approach using a coherent buffer in a nanoscale waveguide by transferring the optical signal coherently to an acoustic wave, and that can be extracted by the reverse process. The life time of the acoustic wave put limitation on the applicability of such approach for single photon signals. We introduce a configuration for achieving a slow signal at the level of single photons without gain or loss. The process utilizes photon-phonon Brillouin interactions involving two counter propagating pump fields. The photon storage is achieved through time delay of significantly slow signal inside nanowires. We address the condition for getting negligible influence due to the scattering off thermal phonons.
\end{abstract}

\maketitle

\section{Introduction}

The manipulation and control of photons and phonons in nanoscale structures becomes of big interest recently \cite{Safavi2019,Wolff2021}. Photons and phonons are among the main players for quantum information processing and communications. The storage of optical signals is a critical process in information technology and implies photonic memory devices \cite{Zhu2007}. Impressive progress has been achieved in optical storage using different techniques with high efficiency performance, e.g. in whispering gallery mode resonators \cite{Dong2015,Kim2015} and optomechanical cavities \cite{Fang2016,Balram2016,Li2015}.  A successful recent approach is in using a coherent buffer in an integrated planar optical waveguide by transferring the optical information coherently to an acoustic wave using Stimulated Brillouin Scattering (SBS) that can be extracted using a reverse process \cite{Merklein2017,Stiller2020,Stiller2024}. The memory device here relies on the transfer from fast optical pulse to slow propagating sound waves. Other schemes that rely on slow light of reduced group velocity have been demonstrated for coupled optical resonators \cite{Xia2007}, photonic crystal cavities \cite{Baba2008,Kuramochi2014}, and optical fibers \cite{Thevenaz2008}.

In recent years, significant progress has been achieved in fabricating waveguides with cross-sections nearing nanoscale dimensions~\cite{Safavi2019}, opening new horizons for SBS. A pivotal advancement in SBS emerged with the identification of a dominant mechanism induced by radiation pressure, as theoretically predicted by \cite{Rakich2012,Rakich2018,VanLaer2016,Zoubi2016} and experimentally realized by \cite{Shin2013,Beugnot2014,VanLaer2015a,VanLaer2015b,Kittlaus2016,Kittlaus2017}. Motivated by such experiments that show significant SBS in
nanoscale waveguides, we investigate the storage of light in such systems \cite{Rakich2012,Shin2013,Kittlaus2016,VanLaer2015a,VanLaer2015b}. The
enhancement of photon-phonon coupling and the long phonon lifetime make
nanoscale waveguides a promising candidate as memory devices. Moreover the
process can be studied on the level of single photons with manifestly quantum
phenomena. The main obstacle here is due to thermal phonons that may forbid
the storage of single photons. We study the condition and limitation for successful
storage of single photons in nanoscale waveguides.

Storing an optical pulse by converting it into long lived acoustic excitations
in an optical fiber through SBS has been presented in \cite{Zhu2007}. The pulse can be retrieved later after a time shorter than the
acoustic excitation lifetime. The method can be used efficiently to store
data coherently for a while. The system can serve as an optical buffer that
temporarily stores light without the need to convert it to electronic signal. A signal pulse can be coherently converted into a mechanical excitation in a
waveguide through SBS involving other counter propagating sharp control
pulse. After a time shorter than the mechanical excitation lifetime the signal
pulse can be retrieved by sending a second control pulse through SBS with the
stored mechanical excitation. Recently coherent phonons have been directly excited by
optical photons through acousto-optic strong coupling in integrated circuit realizing
a coherent all-optical memory \cite{Merklein2017}. The experiment demonstrates a coherent on-chip
memory that allows storing the entire coherent information carried by light as acoustic phonons. We show that such devices work for optical pulses with high average number of photons and fail at pulse with several photons.

In the present paper we emphasize photon number limitation for storage of light and introduce a technique that operates at the level of single photons that relies on slow light configuration. We introduce a configuration for achieving slow photons using SBS within waveguides. By coupling a signal field to classical pump fields through Brillouin scattering mediated by acoustic waves it is possible to achieve a low effective group velocity \cite{Zoubi2017, Zoubi2024}. A stable signal amplitude can be maintained by employing two simultaneous pump fields with frequencies both above and below that of the signal. The Brillouin scattering from the higher pump field into the signal is balanced by the scattering from the signal field into the lower pump field. We account for the impact of thermal phonons in the waveguide medium and identify conditions under which thermal contributions to the signal amplitude are negligible. The signal field, down to the level of single photons, can propagate through the waveguide without any gain or loss, with an effective group velocity significantly reduced to values of the order of the sound velocity.

The paper is organized as follows. In section 2 after introducing the coupling of photons and phonons through Brillouin interactions inside nanowires, we present a configuration for the storage of light by converting the optical signal into an acoustic wave. Section 3 presents a configuration for the storage of light achieved due to slowing light through Brillouin scattering inside nanowires. Conclusions appears in section 4, with emphasize on the possibility of storage of light at the level of single photons for the two previous configurations.

\section{Storage of light}

\begin{figure}[t]
\includegraphics[width=0.8\linewidth]{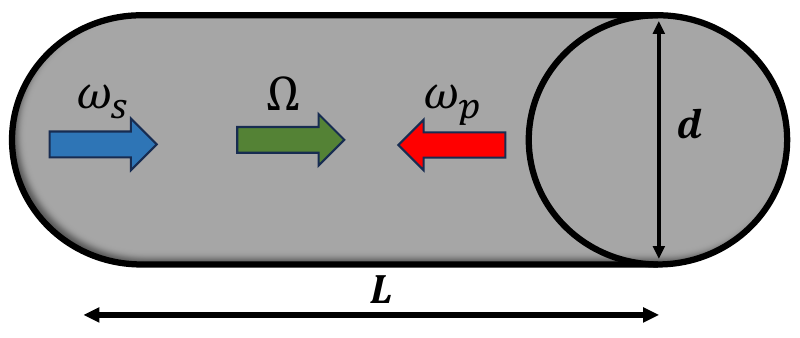}
\caption{A schematic diagram of a waveguide of length $L$ and dimension $d$, where $L\gg d$. The pump and signal fields and the acoustic mode are presented. The light wavelength $\lambda$ obeys $\lambda\lesssim d$.}
 \label{Fig1}
\end{figure}

The memory device treated in the present paper is built of a nanoscale wire composed of dielectric material placed in free space, characterized by a refractive index $n$ greater than one (e.g., for silicon material, $n\approx 3.5$), as depicted in figure (\ref{Fig1}). The length of the waveguide, $L$, significantly exceeds its transverse dimension, $d$, with $L\gg d$, and the light wavelength $\lambda$ is comparable to the wire dimension, $\lambda\lesssim d$. We start by presenting a system of interacting light and sound waves within nanoscale waveguides via Brillouin scattering. In our prior research \cite{Zoubi2016}, we formulated a microscopic quantum theory for the interaction between the light field and mechanical excitations in nanoscale waveguides, deriving a Brillouin-type Hamiltonian for the interplay of photons and phonons. This configuration allows photons and phonons to propagate freely along the waveguide while being confined in the transverse direction, leading to the emergence of photonic and phononic multi-mode branches. In \cite{Zoubi2016} we derived the dispersion relations for photons and phonons and determined the photon-phonon coupling parameter by considering both electrostriction and radiation pressure mechanisms. In such an environment, the coupling of photon-phonon via Brillouin scattering is significantly intensified compared to conventional waveguides, a phenomenon corroborated by experimental findings \cite{Rakich2012,Shin2013,Kharel2016,VanLaer2016,Zoubi2016,Beugnot2014,VanLaer2015a,VanLaer2015b,Kittlaus2016,Kittlaus2017}.

\begin{figure}[t]
\includegraphics[width=0.8\linewidth]{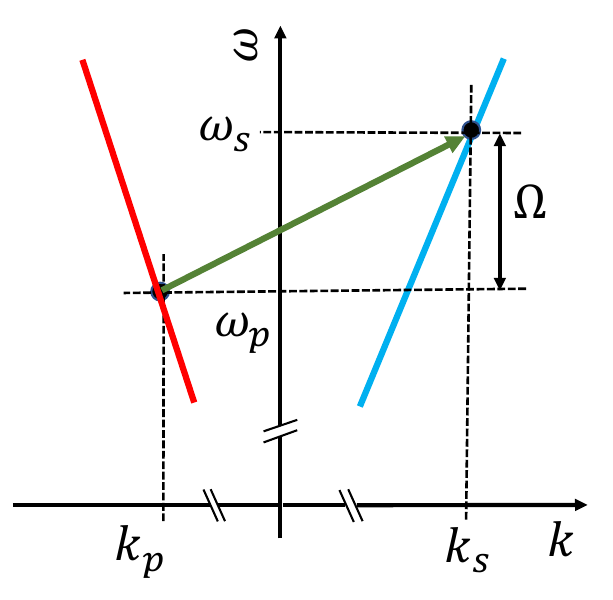}
\caption{The photonic branch is presented for the angular frequency $\omega$ as a function of the wavenumber $k$. The two couterpropagating branches are assumed to have linear dispersion in the appropriate zones with the same group velocity $v_g$. The relevant photon modes treated in the paper are indicated, which are the pump field $(\omega_p,k_p)$, and the signal field $(\omega_s,k_s)$. The phononic mode of the angular frequency $\Omega$ is presented.}
 \label{Fig2}
\end{figure}

In \cite{Zoubi2016}, we solved the equations of motion for the electromagnetic field and mechanical excitation to derive the photon and phonon dispersions analytically for the specific case of a cylindrical waveguide, obtaining the photon frequencies $\omega_{k}$, and the phonon frequency $\Omega_{q}$, where we denote the phonon wave number by $k$, and the phonon wavenumber by $q$. However, the scheme of the current paper can be implemented experimentally for nanoscale wires of any cross-section shape, e.g. circular and rectangular \cite{Rakich2012,Kittlaus2017,Safavi2019}. We focus here on a linear region of the dispersion and assume that the light injected into the waveguide possesses a finite bandwidth. For photons, we employ the linear dispersion relation $\omega_{k} = \omega_{0} + v_{g}(k - k_{0})$, where $\omega_{0}$ is the frequency at the center of the signal bandwidth. The effective group velocity in the linear segment is $v_{g}$, as depicted in figure (\ref{Fig2}). A similar approach is applied to the phonon dispersion, where $\Omega_{q} = \Omega_{0} + v_{s}(q - q_{0})$, with the sound velocity being $v_{s}$. For both propagating photons and phonons, the wavenumbers are determined by the periodic boundary condition in a wire of length $L$, where the wavenumber is quantized as $k = \frac{2\pi}{L}m$ with $m$ being integers $(m = 0, \pm1, \pm2, \cdots)$.

In the following we consider the lowest photonic branch of group velocity $v_g$, and concentrate in modes around the signal frequency $\omega_s$ and the pump frequency $\omega_p$. We consider one of the sound phononic branches of sound velocity $v_s$ and concentrate in modes around the frequency $\Omega$. The real-space Hamiltonian for right-propagating signal field within a waveguide is described by
\begin{equation}
H_{\mathrm{phot}}=\hbar\omega_{s}\int dx\ \hat{\psi}_{s}^{\dagger}(x)\hat{\psi}_{s}(x)-i\hbar v_{g}\int dx\ \hat{\psi}_{s}^{\dagger}(x)\frac{\partial\hat{\psi}_{s}(x)}{\partial x},
\end{equation}
where $\hat{\psi}_{s}^{\dagger}$ and $\hat{\psi}_{s}$ represent the creation and annihilation operators for the signal photon field. The real-space phonon Hamiltonian is given by
\begin{equation}
H_{\mathrm{phon}}=\hbar\Omega\int dx\ \hat{\cal Q}^{\dagger}(x)\hat{\cal Q}(x)-i\hbar v_{s}\int dx\ \hat{\cal Q}^{\dagger}(x)\frac{\partial\hat{\cal Q}(x)}{\partial x},
\end{equation}
where $\hat{\cal Q}^{\dagger}$ and $\hat{\cal Q}$ represent the creation and annihilation operators for the phonon field. Furthermore, the pump pulse is strong and can be treated as a classical field, which is represented by the field amplitude ${\cal E}_p(x)$. The real-space photon and phonon field operators have a dimension of $1/\sqrt{\text{length}}$.

The signal field operator can be described by $\psi_s(x)=\frac{1}{\sqrt{L}}\sum_{k\in{\cal B}}a_ke^{i(k-k_0)x}$, where $a_k$ is a photon annihilation operator at wavenumber $k$, and ${\cal B}$ is the wavenumber bandwidth centered around $k_0$, where the photon dispersion is taken to be linear along such bandwidth. The frequency bandwidth of the signal photon is denoted by $\Delta\tilde{\omega}$ and is much smaller than the phonon frequency, that is $\Delta\tilde{\omega}\ll \Omega$.

The coupling parameter for photon-phonon interaction, $f$, is considered constant across the photon and phonon bandwidths, and we utilize the local field approximation. Consequently, the real-space photon-phonon interaction Hamiltonian is expressed as
\begin{eqnarray}
H_{\mathrm{phot-phon}}&=&i\hbar f\sqrt{L}\int dx\ \left\{\hat{\cal Q}^{\dagger}(x){\cal E}_{p}^{\ast}(x)\hat{\psi}_{s}(x)\right. \nonumber \\
&-&\left.\hat{\psi}_{s}^{\dagger}(x){\cal E}_{p}(x)\hat{\cal Q}(x)\right\}.
\end{eqnarray}
The first term describes the scattering of a signal photon to a pump photon through the emission of a phonon. Conversely, the second term accounts for scattering of a pump photon to a signal photon via the absorption of a phonon. Owing to transnational symmetry along the wire, these processes adhere to momentum conservation, where $k_s-k_p=q$. Moreover, we define the slowly varying fields by
\begin{eqnarray}
{\cal E}_p(x,t)&=&A_p(x,t)\ e^{-i(k_px+\omega_pt)}, \nonumber \\
\hat{\psi}_s(x,t)&=&A_s(x,t)\ e^{i(k_sx-\omega_st)}, \nonumber \\
\hat{\cal Q}(x,t)&=&Q(x,t)\ e^{i(qx-\Omega t)},
\end{eqnarray}
and assume conservation of energy $\omega_s-\omega_p=\Omega$. Furthermore, we consider the photon damping rate $\gamma$ and the phonon damping rate $\Gamma$ phenomenologically, and assume the limit of $\Gamma\gg\gamma$, then we neglect the photon damping during its propagating along the waveguide. Due to the fact that the sound velocity is much smaller than the group velocity $v_g\gg v_s$, one can neglect the phonon propagation along the waveguide and drop the term that includes the position derivative in the phonon Hamiltonian.

The phonon and photon field equations of motion read
\begin{eqnarray}
&&\left(\frac{\partial}{\partial t}+v_g\frac{\partial}{\partial
  x}\right)A_{s}(x,t)=-\sqrt{L}f^{\mu} Q(x,t)A_p(x,t), \nonumber \\
&&\frac{\partial}{\partial t}Q(x,t)=-\frac{\Gamma}{2}\ Q(x,t)+\sqrt{L}f\ A_p^{\dagger}(x,t)A_s(x,t). \nonumber \\
\end{eqnarray}
We neglect changes in the pulse pump field during the interaction, and we apply the change of variables $\xi=x-v_gt,\ \ \ \eta=-x$, with the derivatives $\frac{\partial}{\partial
  x}=\frac{\partial}{\partial\xi}-\frac{\partial}{\partial\eta},\ \ \ \frac{\partial}{\partial
  t}=-v_g\frac{\partial}{\partial\xi}$, and then $\frac{\partial}{\partial t}+v_g\frac{\partial}{\partial
  x}=-v_g\frac{\partial}{\partial\eta}$. The system of equations now read
\begin{eqnarray}
\frac{\partial}{\partial \eta}A_{s}(\eta,\xi)&=&\frac{\sqrt{L}f}{v_g}\ Q(\eta,\xi)A_p(\eta,\xi), \nonumber \\
\frac{\partial}{\partial\xi}Q(\eta,\xi)&=&\frac{\Gamma}{2v_g}\ Q(\eta,\xi)-\frac{\sqrt{L}f}{v_g}\ A_p^{\dagger}(\eta,\xi)A_s(\eta,\xi). \nonumber \\
\end{eqnarray}

\subsection{Storage and Retrieve of Light}

We discuss the storage and retrieve of light in a nanoscale waveguide
stimulated by Brillouin scattering, where the optical signal is converted to an acoustic wave and retrieved later. The initial signal pulse propagates to the right with group velocity $v_g$ and
amplitude $A_s(x,t)=A_s(x-v_gt)=A_s(\xi)$ of duration
$\Lambda$. The
control pulse propagates to the left with group velocity $v_g$ and
amplitude $A_p(x,t)=A_p(x+v_gt)=A_p(-\xi-2\eta)$ of duration
$\sigma$. The control pulse duration is taken to be much smaller than the signal pulse
one, that is $\Lambda\gg\sigma$.

For illustration we consider first the case of
a very sharp control pulse that can be represented by a delta function, which we
term as a write pulse, where $A_p^w(-\xi-2\eta)={\cal E}\delta(-\xi-2\eta)$. In
neglecting the phonon damping during the signal pulse duration we get
\begin{eqnarray}
Q(\xi,\eta)&=&-\frac{\sqrt{L}f}{v_g}{\cal E}^{\ast}\int d\xi'\ \delta(-\xi'-2\eta)A_s(\xi') \nonumber \\
&=&-\frac{\sqrt{L}f}{v_g}{\cal E}^{\ast}A_s(-2\eta),
\end{eqnarray}
or
\begin{equation}
Q(x,t)=-\frac{\sqrt{L}f}{v_g}{\cal E}^{\ast}A_s(2x).
\end{equation}
The signal pulse is converted into a mechanical excitation with half size of
the signal.

After time $\Delta$ we send a second control pulse, a read pulse, of the shape
$A_p^r(\xi)={\cal E}\delta(-\xi-2\eta-v_g\Delta)$. During time $\Delta$ the
acoustic excitation decays with a damping rate $\Gamma$. Hence just before the
read pulse meets the acoustic excitation we have
\begin{equation}
Q(\eta,\xi)=-\frac{\sqrt{L}f}{v_g}{\cal
  E}^{\ast}A_s(-2\eta)\ e^{-\Gamma\Delta/2}.
\end{equation}
Therefore to have an efficient write and read of a signal one needs to keep
$\Delta<1/\Gamma$. We obtain
\begin{eqnarray}
A_{s}(\eta,\xi)&=&-\frac{Lf^2|{\cal E}|^2}{v_g^2}\ e^{-\Gamma\Delta/2} \nonumber \\
&\times&\int d\eta'\delta(-\xi-2\eta'-v_g\Delta)A_s(-2\eta'),
\end{eqnarray}
that yields
\begin{equation}
A_{s}(\eta,\xi)=-\frac{Lf^2|{\cal E}|^2}{v_g^2}\ e^{-\Gamma\Delta/2}\ A_s(\xi+v_g\Delta),
\end{equation}
or
\begin{equation}
A_{s}(x,t)=-\frac{Lf^2|{\cal E}|^2}{v_g^2}\ e^{-\Gamma\Delta/2}\ A_s(x-v_gt+v_g\Delta),
\end{equation}
We conclude that the signal pulse was coherently stored as a mechanical excitation for a
time $\Delta$ that is shorter than the acoustic excitation lifetime and
coherently retrieved afterward with a reduced amplitude.

Next we consider the
case of a write control pulse $A_{p}^w(-\xi-2\eta)$ with a finite duration
$\sigma$, and a signal pulse $A_{s}^{in}(\xi)$ of duration
$\Lambda$. We take $\sigma$ to be much smaller
than $\Lambda$, that is $\Lambda\gg\sigma$. We consider the lowest order
perturbative solution where the field amplitudes are taken to be fixed during
the scattering time. Directly after the control pulse passed the signal pulse,
the solution for the signal pulse reads (see, e.g., \cite{Dodin2002})
\begin{equation}
A_{s}(\eta,\xi)=A_{s}^{in}(\xi)\cos\Theta_w,
\end{equation}
where the area function is defined as
\begin{equation}
\Theta_w=\frac{\sqrt{L}f}{v_g}\int_{0}^{-\xi-\eta}d\zeta\ A_{p}^w(\zeta).
\end{equation}
For the mechanical excitation we get
\begin{equation}
Q(\eta,\xi)=A_{s}^{in}(-2\eta)\sin\Theta_w,
\end{equation}
which is a localized excitation with half the size of the original signal
pulse. After time $\Delta$, due to acoustic excitation damping, we get
\begin{equation}
Q(\eta,\xi)=A_{s}^{in}(-2\eta)\sin\Theta_w\ e^{-\Gamma\Delta/2}.
\end{equation}
We send a second read control pulse $A_{p}^r(-\xi-2\eta-v_g\Delta)$ that meet the mechanical excitation after time
$\Delta$, which is smaller than the phonon lifetime, that is $\Delta<1/\Gamma$. In the end of the overlap between the control pulse and the
mechanical excitation we have
\begin{equation}
Q(\eta,\xi)=A_{s}^{in}(-2\eta)\sin\Theta_w\cos\Theta_r\ e^{-\Gamma\Delta/2},
\end{equation}
and the output signal field reads
\begin{equation}
A_{s}^{out}(\eta,\xi)=A_{s}^{in}(\xi+v_g\Delta)\sin\Theta_w\sin\Theta_r\ e^{-\Gamma\Delta/2},
\end{equation}
where the area function is defined by
\begin{equation}
\Theta_r=\frac{\sqrt{L}f}{v_g}\int_{0}^{-\xi-\eta}d\zeta\ A_{p}^r(\zeta).
\end{equation}

The solution can be also represented in $(x,t)$ variables, for an input
signal pulse $A_s(x,t)=A_s^{(in)}(x-v_gt)$ and a control pulse
$A_p(x,t)=A_p(x+v_gt)$. After sending the
first control pulse, we have
\begin{equation}
A_{s}(x-v_gt)=A_{s}^{in}(x-v_gt)\cos\Theta_w,
\end{equation}
and
\begin{equation}
Q(x,t)=A_{s}^{in}(2x)\sin\Theta_w,
\end{equation}
where
\begin{equation}
\Theta_w=\sqrt{L}f\int_{0}^{t}dt'\ A_{p}^w(t').
\end{equation}
After time $\Delta$, we have
\begin{equation}
Q=A_{s}^{in}(2x)\sin\Theta_w\ e^{-\Gamma\Delta/2}.
\end{equation}
After sending the second control pulse, we get
\begin{equation}
Q(x,t)=A_{s}^{in}(2x)\sin\Theta_w\cos\Theta_r\ e^{-\Gamma\Delta/2},
\end{equation}
and
\begin{equation}
A_{s}^{out}(x,t)=A_{s}^{in}(x-v_g(t-\Delta))\sin\Theta_w\sin\Theta_r\ e^{-\Gamma\Delta/2},
\end{equation}
with
\begin{equation}
\Theta_r=\sqrt{L}f\int_{0}^{t}dt'\ A_{p}^r(t').
\end{equation}

Complete storage and retrieval are obtained for $\Theta_w=\Theta_r=(n+1/2)\pi$,
where $n=0,1,2,\cdots$. For example, at $\Theta=\pi/2$ we get after the first
control pulse $A_{s}(x-v_gt)=0$ and
$Q(x,t)=A_{s}^{in}(2x)$. After the second control
pulse we have $Q(x,t)=0$ and
$A_{s}^{out}(x,t)=A_{s}^{in}(x-v_g(t-\Delta))\ e^{-\Gamma\Delta/2}$. The
area function includes the control pulse amplitude and the coupling parameter,
the fact that gives us a flexibility for getting  the required values for the
area parameters. The key parameter in the above process is the lifetime of
the phonons.

\begin{figure}
\includegraphics[width=0.8\linewidth]{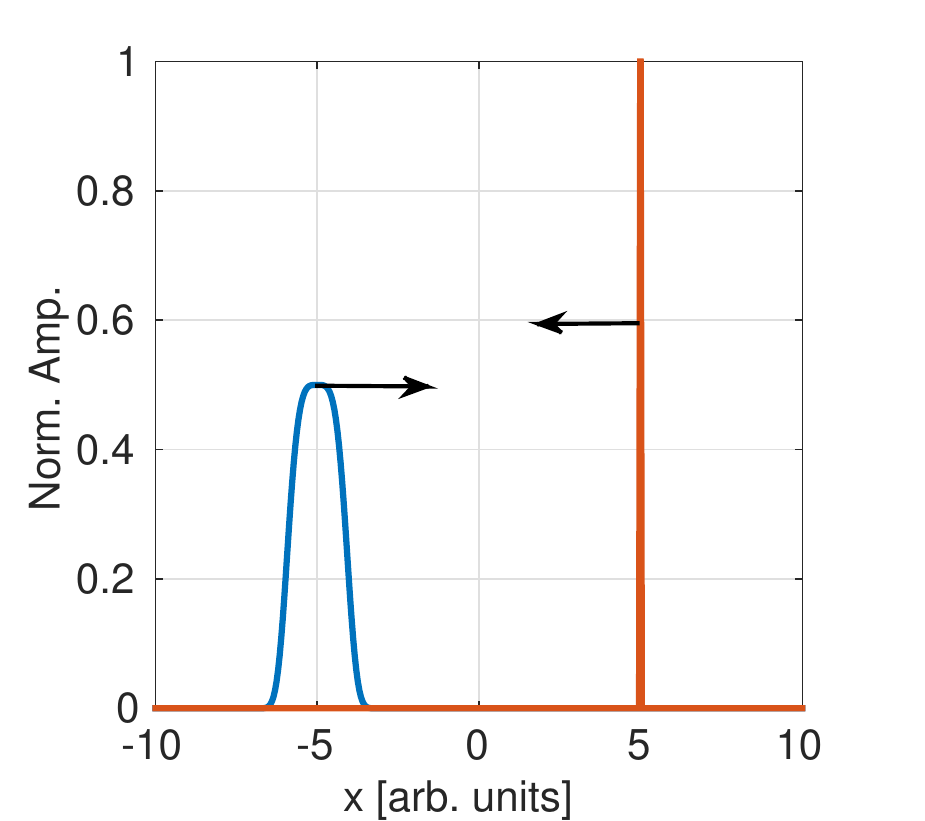}
\caption{At $t=-5\ ns$: An incident signal pulse (blue) propagates to the
  right of pulse duration $\Lambda=10\ ns$. A counter propagating
  control write pulse (red) of pulse duration $\sigma=1\ ns$
  propagates to the left.}
\label{Pulse1}
\end{figure}

\begin{figure}
\includegraphics[width=0.8\linewidth]{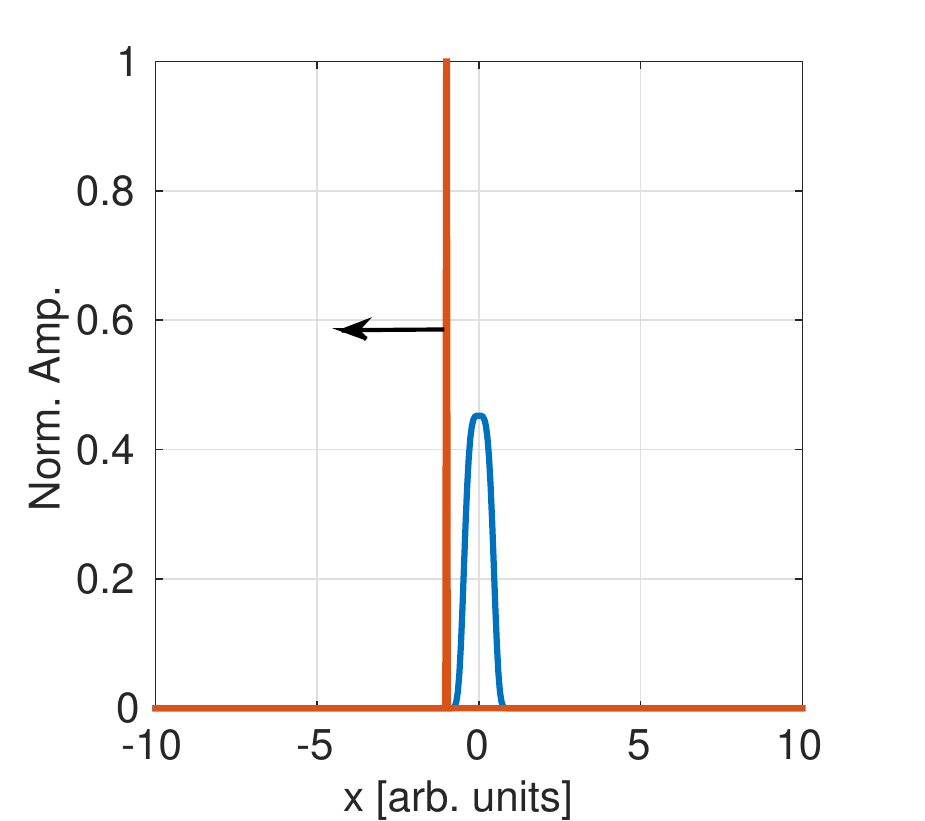}
\caption{At $t=1\ ns$: The control write pulse (red) continue to the left. A
  localized acoustic excitation (blue) is
  produced of half of the signal size.}
\label{Pulse2}
\end{figure}

\begin{figure}
\includegraphics[width=0.8\linewidth]{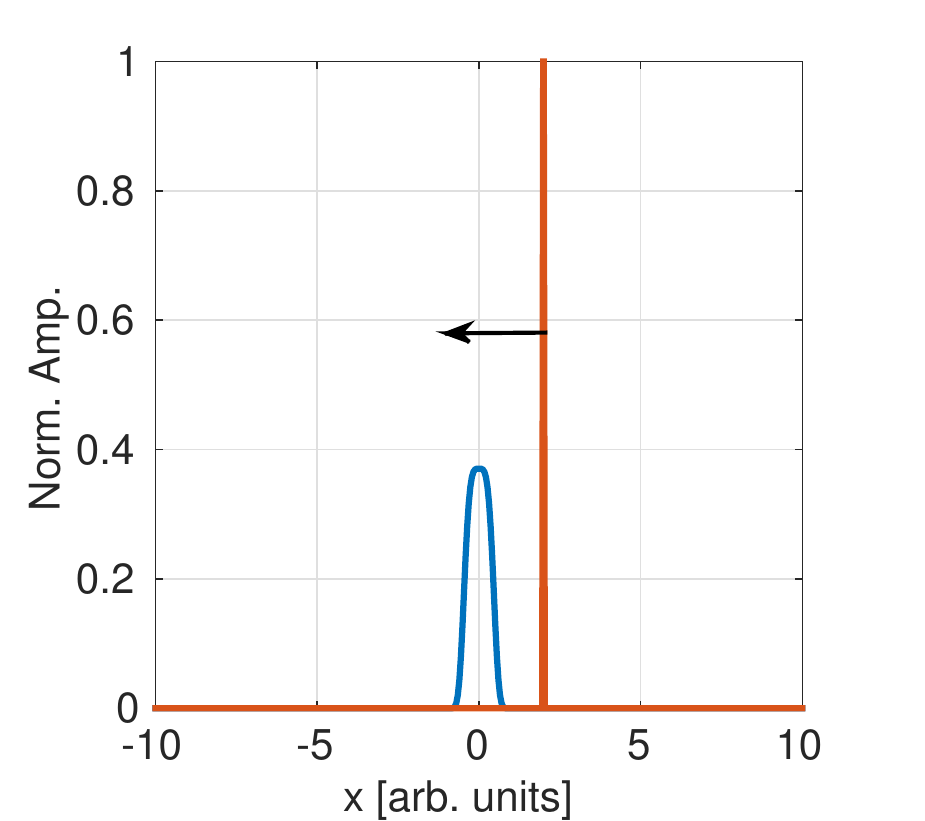}
\caption{At $t=3\ ns$: After time $\Delta=5\ ns$ a second read control pulse
  (red) meets the acoustic excitation (blue) from the right. At
  time $\Delta$ the acoustic excitation amplitude is reduced by a factor
  $e^{-\Gamma\Delta/2}$.}
\label{Pulse3}
\end{figure}

\begin{figure}
\includegraphics[width=0.8\linewidth]{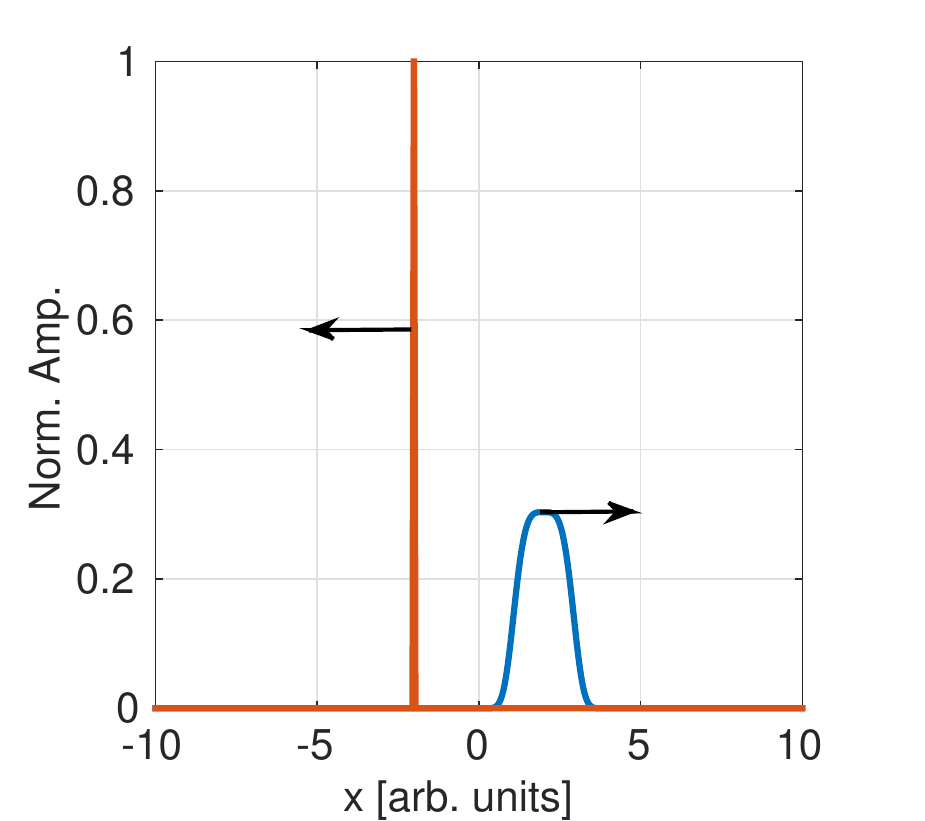}
\caption{At $t=7\ ns$: The
read pulse (red) continue to propagate to the left. The signal pulse (blue) is retrieved
and propagates to the right with reduced
amplitude. The retrieved pulse is delayed by $\Delta$ from the original
signal pulse.}
\label{Pulse4}
\end{figure}

Quantitative plots are represented in figures (\ref{Pulse1}-\ref{Pulse4}). We
use a control pulse of duration $\sigma=1\ ns$ and a signal pulse of duration
$\Lambda=10\ ns$. The pulses are plotted using the function $A\propto
e^{-(x\pm v_gt)^{2r}}$, where $r$ is an integer, with $r=2$ for the signal pulse and $r=10$ for the
control pulse. Figure (\ref{Pulse1}) is at $t=-5\ ns$, where the control pulse
propagates to the left and the signal pulse to the right. After the interaction
the signal disappears and a mechanical excitation appears, as seen in figure
(\ref{Pulse2}) at $t=1\ ns$, where the control pulse continue to the
left. After waiting more time at $t=3\ ns$ the excitation amplitude damped,
as seen in figure (\ref{Pulse3}), with damping rate of $1/\Gamma=10\ ns$. The
second control read pulse is sent
from the right. Finally, in plot (\ref{Pulse4}) at $t=7\ ns$ the signal pulse retrieved and
move to the right with time delay of $\Delta=5\ ns$.

The current scenario is useful for a signal field that includes relatively large number of photons, and useless for a dilute signal that contains several photons, where the phonon dissipation during the storage period put limitation on the efficiency of the current configuration. In the next section we introduce a storage process that rest on a signal time delay due to slow photons, where the suggested configuration works efficiently for a signal of single photons. But first let us consider the influence of thermal phonon on the storage process treated above.

\subsection{Thermal Phonon}

In the previous treatment the effect of thermal phonons was neglected, where we
included only phenomenologically the damping of the signal phonon during the
arrival of the
two pump pulses. Here we present more detail treatment of the influence of
thermal phonons in using quantum Langevin method \cite{Boyd1990,Zoubi2017}. The pulse duration $\sigma$
is much shorter than the signal duration $\Lambda$, that is
$\sigma\ll\Lambda$, the fact that allows us to neglect the effect of thermal
phonons during the Brillouin interaction among the two light fields. Our main
concern here is on the production of signal photons due to the scattering of
pump photons from thermal phonons during the propagation of the pump pulse
along the waveguide. As we deal with a signal of few photons then we are
looking for the condition in which the emitted signal photons out of this process
is much smaller than one, or exactly $N_s\gg N_f$, where $N_s=L\langle
A_s^{\dagger}A_s\rangle$ is the average number of signal photons, and $N_f$ is
the average number of scattered photons that induced by thermal fluctuations.

We concentrate on the scattering of pump photons into signal photons involving
thermal phonons, we solve the system of equations
\begin{eqnarray}
\frac{\partial}{\partial \eta}A_{s}(\eta,\xi)&=&\frac{\sqrt{L}f}{v_g}\ Q(\eta,\xi)A_p(\eta,\xi), \nonumber \\
\frac{\partial}{\partial\xi}Q(\eta,\xi)&=&\frac{\Gamma}{2v_g}\ Q(\eta,\xi)-{\cal
  F}(\eta,\xi), \nonumber \\
\end{eqnarray}
where ${\cal F}(\eta,\xi)$ is the Langevin noise operator. The pump field is treated
classically. The phonon operator is solved formally to give
\begin{equation}
Q(\eta,\xi)=-\int_{0}^{\xi} d\xi'\ {\cal F}(\eta,\xi')e^{\frac{\Gamma}{2v_g}(\xi'-\xi)}.
\end{equation}
Substituting in the signal field equation, yields
\begin{equation}
\frac{\partial}{\partial \eta}A_{s}(\eta,\xi)=-\frac{\sqrt{L}f}{v_g}A_p(\eta,\xi)\int_{0}^{\xi} d\xi'\ {\cal F}(\eta,\xi')e^{\frac{\Gamma}{2v_g}(\xi'-\xi)},
\end{equation}
and formal integration gives
\begin{equation}
A_{s}(\eta,\xi)=-\frac{\sqrt{L}f}{v_g}\int_{0}^{\eta} d\eta'\int_{0}^{\xi} d\xi'A_p(\eta',\xi){\cal F}(\eta',\xi')e^{\frac{\Gamma}{2v_g}(\xi'-\xi)}.
\end{equation}
The average number of fluctuation signal photons is
\begin{eqnarray}
N_f&=&\frac{L^2|f|^2}{v_g^2}\int_{0}^{\eta}
d\eta'd\eta''\int_{0}^{\xi}
d\xi'd\xi''\ A_p^{\ast}(\eta',\xi)A_p(\eta'',\xi) \nonumber \\
&\times&\langle{\cal
  F}^{\dagger}(\eta',\xi'){\cal F}(\eta'',\xi'')\rangle\ e^{\frac{\Gamma}{2v_g}\left[(\xi'-\xi)+(\xi''-\xi)\right]}.
\end{eqnarray}
We use the phonon reservoir property
\begin{equation}
\langle{\cal
  F}^{\dagger}(\eta',\xi'){\cal F}(\eta'',\xi'')\rangle=\frac{\Gamma \bar{n}_0}{v_g}\delta(\eta'-\eta'')\delta(\xi'-\xi''),
\end{equation}
where $\bar{n}_0$ the average number of thermal phonons, to obtain
\begin{equation}
N_f=\frac{L^2|f|^2}{v_g^2}\frac{\Gamma \bar{n}_0}{v_g}\int_{0}^{\eta}
d\eta'\int_{0}^{\xi}
d\xi'\ |A_p(\eta',\xi)|^2\ e^{\frac{\Gamma}{v_g}(\xi'-\xi)}.
\end{equation}
Integration over $\xi'$, yields
\begin{equation}
N_f=\frac{L^3|f|^2}{v_g^3}\Gamma\bar{n}_0\int_{0}^{\eta}
d\eta'\ |A_p(\eta',\xi)|^2,
\end{equation}
where $\Gamma L/v_g\ll 1$. We have
\begin{equation}
N_f=\frac{L^3|f|^2}{v_g^3}\Gamma\bar{n}_0|{\cal E}|^2.
\end{equation}
where $|{\cal E}|^2=L|A_p|^2$ is the average number of pump photons. For a single photon signal the condition for negligible thermal fluctuation
is $N_f\ll 1$.

To have efficient storage of photons we imply $\Theta=\pi/2$, where
$\Theta\approx|{\cal E}||f|\sigma$. For a pump pulse
of duration $\sigma=0.1$~ns we need $|f||{\cal
  E}|=\frac{\pi}{2}10^{10}$~Hz. Here, for $\Gamma = 1\ MHz$, we have
$N_f\approx\bar{n}_0$, for $L=10$~cm and $v_g\approx c/5$, where $\Gamma L/v_g\approx 10^{-3}$. For phonon frequency of $\Omega=10$~GHz, we have $Q=10^4$, and the temperature
$T=200$~mK, we get $\bar{n}_0\approx0.1$.

\section{Slow Light}

Now we present  a configuration for the storage of light by obtaining slow light inside nanowires. Our primary interest here lies in slowing down the group velocity of optical signal at the level of single photons. Our objective is to attain a propagating signal with an effective group velocity significantly lower than that in free space, while also ensuring a constant average number of quanta. Additionally, it's crucial to minimize the impact of thermal fluctuations, preventing them from significantly affecting the propagating signal. Therefore, our goal is to introduce a configuration that enables the realization of slow signals at the single-photon level without inducing gain or loss.

\begin{figure}[t]
\includegraphics[width=0.8\linewidth]{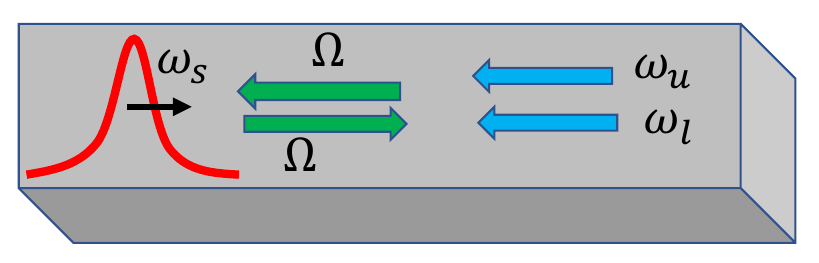}
\caption{A signal field of frequency $\omega_s$ is propagating to the right, with two counter-propagating classical pump fields of frequencies $\omega_u$ and $\omega_l$, where $\omega_u>\omega_s>\omega_l$. Due to stimulated Brillouin scattering a signal photon scatters into a pump photon of frequency $\omega_l$ by the emission of a co-propagating phonon of frequency $\Omega$, and a pump photon of frequency $\omega_u$ scatters into a signal photon by the emission of a counter-propagating phonon of the same frequency.}
 \label{Fig7}
\end{figure}

\begin{figure}[t]
\includegraphics[width=0.8\linewidth]{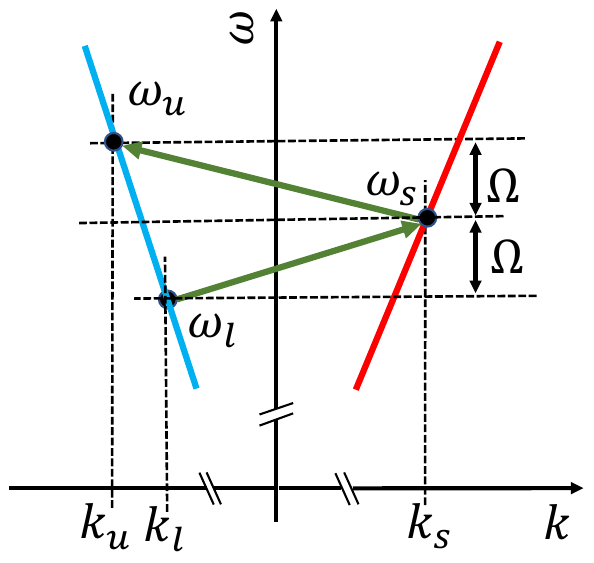}
\caption{A pump field $(\omega_u)$ scatters into a signal field $(\omega_s)$ by the emission of a phonon $(\Omega)$, and a signal field scatters into a pump field $(\omega_l)$ by the emission of a phonon of the same frequency. The two phonon differs in their wavenumbers.}
 \label{Fig8}
\end{figure}

\begin{figure}[t]
\includegraphics[width=0.8\linewidth]{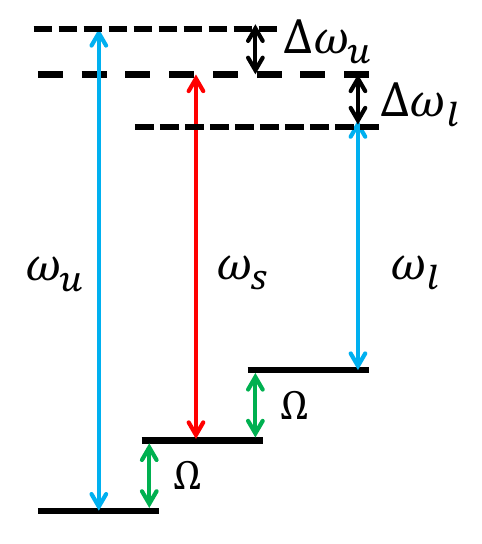}
\caption{A schematic energy diagram of the photon and phonon modes for the two processes. A pump photon (of frequency $\omega_u$) is annihilated and a signal photon (of frequency $\omega_s$) and a phonon (of frequency $\Omega$) are created, with the detuning frequency $\Delta\omega_u=\omega_u-\omega_s-\Omega$. A signal photon is annihilated and a pump photon (of frequency $\omega_l$) and a phonon (of the same frequency) are created, with the detuning frequency $\Delta\omega_l=\omega_s-\omega_l-\Omega$.}
 \label{Fig9}
\end{figure}

To address the challenges, we propose a unique configuration in which the signal field is coupled through SBS to two counter propagating pump fields, involving a dispersion-less vibration mode. This approach aims to demonstrate that a slow signal can be achieved without gain or loss. Specifically, a signal field in the lowest branch centered around frequency $\omega_s$ and group velocity $v_g$ is coupled to two counter propagating pump fields with amplitudes ${\cal E}_l$ and ${\cal E}_u$, and frequencies $\omega_l$ and $\omega_u$ respectively, where $\omega_u>\omega_s>\omega_l$, as depicted in figures (\ref{Fig7}-\ref{Fig8}). Here the pump fields, being considerably stronger than the signal field, are treated as a classical quantity with a stationary (slowly varying) amplitude denoted by ${\cal E}=\langle\hat{\psi}_p\rangle$.  On the phonon side, we assume a non-dispersive single branch with a constant frequency $\Omega$, and negligible sound velocity $v_{s}$. The SBS process adheres to the phase matching condition for coupling with both the upper and lower pump fields. The photon-phonon coupling parameter is considered to be real, local (i.e., wavenumber independent), and identical for both interactions, with $g=g_l=g_u$. Additionally, the lower and upper detuning frequencies are defined as $\Delta\omega_l=\omega_s-\omega_l-\Omega$ and $\Delta\omega_u=\omega_u-\omega_s-\Omega$, respectively, as schematically illustrated in figure (\ref{Fig9}). Both the upper and lower SBS processes involve phonons at the same frequency $\Omega$ but with distinct wavenumbers. The phonon damping rate is denoted by $\Gamma$, and the Langevin force operator $\hat{\cal F}$ is considered identical for both Brillouin scattering processes.

Other nonlinear optical processes, e.g. Kerr scattering, involving the signal and pump fields can be excluded due to phase-matching conditions, as they do not obey conservation of both energy and momentum simultaneously. Moreover conventional nonlinear optics is negligible on the level of single photons and can be omitted, and in the following we concentrate only in the SBS process.

The signal photon operator is  $\hat{\psi}_{s}$, and the photon Hamiltonian is formulated as
\begin{equation}\label{SigHam}
H_{\mathrm{phot}}=\hbar\omega_{s}\int dx\ \hat{\psi}_{s}^{\dagger}(x)\hat{\psi}_{s}(x)-i\hbar v_{g}\int dx\ \hat{\psi}_{s}^{\dagger}(x)\frac{\partial\hat{\psi}_{s}(x)}{\partial x}.
\end{equation}
The phonon operators are expressed by $\hat{\cal Q}_{u}$ and $\hat{\cal Q}_{l}$. The associated Hamiltonian for the phonons is formulated as
\begin{equation}\label{phonu}
H_{\mathrm{phon}}=\hbar\Omega \left(\int dx\ \hat{\cal Q}_{u}^{\dagger}(x)\hat{\cal Q}_{u}(x)+\int dx\ \hat{\cal Q}_{l}^{\dagger}(x)\hat{\cal Q}_{l}(x)\right).
\end{equation}
The interaction Hamiltonian between photons and phonons is given by
\begin{eqnarray}\label{photphonu}
H_{\mathrm{phot-phon}}&=&\hbar\sqrt{L}\int dx\ \left\{g_{u}^{\ast}{\cal E}_u\ \hat{\cal Q}_{u}^{\dagger}(x)\hat{\psi}_{s}^{\dagger}(x)\right. \nonumber \\
&+&\left.g_{l}^{\ast}{\cal E}_l\ \hat{\cal Q}_{l}(x)\hat{\psi}_{s}^{\dagger}(x)+\mathrm{h.c.}
\right\}.
\end{eqnarray}

The Heisenberg-Langevin equation of motion for the photon operator is expressed as
\begin{eqnarray}
\left(\frac{\partial}{\partial t}+v_g\frac{\partial}{\partial x}\right)\hat{\psi}_{s}(x,t)&=&-i\sqrt{L}g{\cal E}_u e^{-i\Delta\omega_ut}\ \hat{\cal Q}_{u}^{\dagger}(x,t) \nonumber \\
&-&i\sqrt{L}g{\cal E}_l e^{i\Delta\omega_lt}\ \hat{\cal Q}_{l}(x,t),
\end{eqnarray}
and the equations of motion for the phonon field operators are given by
\begin{eqnarray}
\left(\frac{\partial}{\partial t}+\frac{\Gamma}{2}\right)\hat{\cal Q}_{u}(x,t)&=&-i\sqrt{L}g_{u}^{\ast}{\cal E}_u e^{-i\Delta\omega_ut}\ \hat{\psi}_{s}^{\dagger}(x,t)-\hat{\cal F}(x,t), \nonumber \\
\left(\frac{\partial}{\partial t}+\frac{\Gamma}{2}\right)\hat{\cal Q}_{l}(x,t)&=&-i\sqrt{L}g_{l}{\cal E}_l^{\ast} e^{-i\Delta\omega_lt}\ \hat{\psi}_{s}(x,t)-\hat{\cal F}(x,t), \nonumber \\
\end{eqnarray}
where we performed the replacements: $\hat{\psi}_{s}(x,t)\rightarrow e^{-i\omega_st}\hat{\psi}_{s}(x,t)$, $\hat{\cal Q}_{u}(x,t)\rightarrow e^{-i\Omega t}\hat{\cal Q}_{u}(x,t)$, $\hat{\cal Q}_{l}(x,t)\rightarrow e^{-i\Omega t}\hat{\cal Q}_{l}(x,t)$, ${\cal E}_u\rightarrow e^{-i\omega_ut}{\cal E}_u$, ${\cal E}_l\rightarrow e^{-i\omega_lt}{\cal E}_l$, and $\hat{\cal F}(x,t)\rightarrow e^{-i\Omega t}\hat{\cal F}(x,t)$.

The rate of photon damping is considered negligible during their transit along the waveguide's length $L$. Phonon dissipation is accounted for by incorporating a damping rate $\Gamma$, and thermal fluctuations are represented through the Langevin force operators $\hat{\cal F}$, adhering to the properties outlined in \cite{Gardiner2010}
\begin{eqnarray}\label{LF}
\langle\hat{\cal F}(x,t)\hat{\cal F}(x',t')\rangle&=&\langle\hat{\cal F}^{\dagger}(x,t)\hat{\cal F}^{\dagger}(x',t')\rangle=0, \nonumber \\
\langle\hat{\cal F}^{\dagger}(x,t)\hat{\cal F}(x',t')\rangle&=&\Gamma \bar{n}\ \delta(t-t')\delta(x-x'), \nonumber \\
\langle\hat{\cal F}(x,t)\hat{\cal F}^{\dagger}(x',t')\rangle&=&\Gamma (\bar{n}+1)\ \delta(t-t')\delta(x-x'),
\end{eqnarray}
with $\bar{n}$ representing the average phonon count at frequency $\Omega$. At low temperatures the appearance of thermal photons is negligible, while thermal phonons are likely present and treated here as a heat reservoir in applying the Markovian approximation \cite{Gardiner2010}.

Formal integration of the phonon operator equations gives
\begin{align}
\hat{\cal Q}_{u}(x,t)&=\hat{\cal Q}_{u}(x,0)e^{-\Gamma t/2}\nonumber\\
&\quad-i\sqrt{L}g_{u}^{\ast}{\cal E}_u\int_0^tdt'\ \hat{\psi}_{s}^{\dagger}(x,t')\ e^{-i\Delta\omega_ut'}e^{-\frac{\Gamma}{2}(t-t')}\nonumber\\
&-\int_0^tdt'\ \hat{\cal F}(x,t')e^{-\frac{\Gamma}{2}(t-t')},\nonumber\\
\hat{\cal Q}_{l}(x,t)&=\hat{\cal Q}_{l}(x,0)e^{-\Gamma t/2}\nonumber\\
&\quad-i\sqrt{L}g_{l}{\cal E}_l^{\ast}\int_0^tdt'\ \hat{\psi}_{s}(x,t')\ e^{-i\Delta\omega_lt'}e^{-\frac{\Gamma}{2}(t-t')}\nonumber\\
&-\int_0^tdt'\ \hat{\cal F}(x,t')e^{-\frac{\Gamma}{2}(t-t')}.
\end{align}
Iterative solution in term of the photon-phonon coupling parameter allows taking the signal operator out of the integral. In neglecting the phonon operator at initial time and after time integration, we get
\begin{align}
\hat{\cal Q}_{u}(x,t)&=-i\sqrt{L}g_{u}^{\ast}{\cal E}_u\frac{e^{-i\Delta\omega_ut}}{\Gamma/2-i\Delta\omega_u}\hat{\psi}_{s}^{\dagger}(x,t)\nonumber\\
&-\int_0^tdt'\ \hat{\cal F}(x,t')e^{-\frac{\Gamma}{2}(t-t')},\nonumber\\
\hat{\cal Q}_{l}(x,t)&=-i\sqrt{L}g_{l}{\cal E}_l^{\ast}\frac{e^{-i\Delta\omega_lt}}{\Gamma/2-i\Delta\omega_l}\hat{\psi}_{s}(x,t)\nonumber\\
&-\int_0^tdt'\ \hat{\cal F}(x,t')e^{-\frac{\Gamma}{2}(t-t')}.
\end{align}

Substituting the phonon operator into the signal operator equation of motion gives
\begin{align}
&\left(\frac{\partial}{\partial t}+v_g\frac{\partial}{\partial x}\right)\hat{\psi}_{s}(x,t)=v_g\left[G_u-G_l-i\kappa_u-i\kappa_l\right]\hat{\psi}_{s}(x,t)\nonumber\\
&\quad+i\sqrt{L}g_{u}^{\ast}{\cal E}_u \hat{\cal W}_u^{\dagger}(x,t)+i\sqrt{L}g_{l}^{\ast}{\cal E}_l \hat{\cal W}_l(x,t),
\end{align}
where the gain parameters are
\begin{equation}\label{Gu}
G_u=\frac{\Gamma|g_u|^2L|{\cal E}_u|^2}{2v_g\left(\Gamma^2/4+\Delta\omega_u^2\right)},\ \ \ G_l=\frac{\Gamma|g_l|^2L|{\cal E}_l|^2}{2v_g\left(\Gamma^2/4+\Delta\omega_l^2\right)},
\end{equation}
and the wavenumber shifts are
\begin{equation}\label{ku}
\kappa_u=\frac{\Delta\omega_u|g_u|^2L|{\cal E}_u|^2}{v_g\left(\Gamma^2/4+\Delta\omega_u^2\right)},\ \ \ \kappa_l=\frac{\Delta\omega_l|g_l|^2L|{\cal E}_l|^2}{v_g\left(\Gamma^2/4+\Delta\omega_l^2\right)}.
\end{equation}
We defined
\begin{eqnarray}\label{Wu}
\hat{\cal W}_{u}(x,t)&=&e^{i\Delta\omega_{u}t}\int_0^tdt'\ \hat{\cal F}(x,t')e^{-\frac{\Gamma}{2}(t-t')}, \nonumber \\
\hat{\cal W}_l(x,t)&=&e^{i\Delta\omega_lt}\int_0^tdt'\ \hat{\cal F}(x,t')e^{-\frac{\Gamma}{2}(t-t')}.
\end{eqnarray}

Solving the equation, we arrive at
\begin{eqnarray}
&&\hat{\psi}_{s}(x,t)=\hat{\psi}_{s}^{in}(x-v_gt)e^{(G-i\kappa)x} \nonumber \\ 
&+&i\frac{\sqrt{L}g}{v_g}\int_0^t dt'\int_0^{x}dx'e^{(G-i\kappa)(x-x')}e^{-\frac{\Gamma}{2}(t-t')} \nonumber \\ 
&\times&\left\{{\cal E}_l\hat{\cal F}(x',t')e^{i\Delta\omega_lt}+{\cal E}_u\hat{\cal F}^{\dagger}(x',t')e^{-i\Delta\omega_ut}\right\}
\end{eqnarray}
where $G=G_u-G_l$ and $\kappa=\kappa_u+\kappa_l$, integrating $G_u,\ \kappa_u$, and $G_l,\ \kappa_l$. The gain $G$ and phase shift $\kappa$ are given by
\begin{equation}
G=\frac{2g^2}{v_g\Gamma}\left\{\frac{{\cal I}_u}{1+\Delta_u^2}-\frac{{\cal I}_l}{1+\Delta_l^2}\right\},
\end{equation}
and
\begin{equation}
\kappa=\frac{2g^2}{v_g\Gamma}\left\{\frac{\Delta_u{\cal I}_u}{1+\Delta_u^2}+\frac{\Delta_l{\cal I}_l}{1+\Delta_l^2} \right\}.
\end{equation}
The key control parameters remain the scaled detunings $\Delta_u=2\Delta\omega_u/\Gamma$ and $\Delta_l=2\Delta\omega_l/\Gamma$, alongside the dimensionless pump intensities ${\cal I}_u=L|{\cal E}_u|^2$ and ${\cal I}_l=L|{\cal E}_l|^2$.

For the photon density, we obtain
\begin{equation}
\langle\hat{\psi}_{s}^{\dagger}(x,t)\hat{\psi}_{s}(x,t)\rangle=\langle\hat{\psi}_{s}^{in\dagger}(x-v_gt)\hat{\psi}_{s}^{in}(x-v_gt)\rangle e^{2Gx}+{\cal N}(x,t),
\end{equation}
where the thermal fluctuation contribution is given by
\begin{equation}
{\cal N}(x,t)=-\frac{g^2}{2Gv_g^2}\left\{{\cal I}_l\bar{n}+{\cal I}_u(\bar{n}+1)\right\}\left(1-e^{-\Gamma t}\right)\left(1-e^{2Gx}\right).
\end{equation}
Utilizing relations (\ref{LF}) for both the upper and lower processes, correlations among the Langevin force operators corresponding to the upper and lower processes are neglected.

The effective group velocity is defined by
\begin{equation}
\frac{1}{v_e}=\frac{1}{v_g}-\frac{\partial\kappa}{\partial\omega_s}.
\end{equation}
We have
\begin{equation}\label{ve}
\frac{v_e}{v_g}=\left(1+\frac{4g^2}{\Gamma^2}\left\{{\cal I}_u\frac{\left[1-\Delta_u^2\right]}{\left[1+\Delta_u^2\right]^2}-{\cal I}_l\frac{\left[1-\Delta_l^2\right]}{\left[1+\Delta_l^2\right]^2}\right\}\right)^{-1}.
\end{equation}
The rate of change of gain with respect to the signal frequency is expressed as
\begin{equation}\label{dG}
\frac{\partial G}{\partial\omega_s}=\frac{8g^2}{v_g\Gamma^2}\left\{{\cal I}_u\frac{\Delta_u}{\left[1+\Delta_u^2\right]^2}+{\cal I}_l\frac{\Delta_l}{\left[1+\Delta_l^2\right]^2}\right\},
\end{equation}
The objective is to achieve a slow propagating signal, where $\frac{v_e}{v_g}\ll1$. Additionally, it is essential for the signal to propagate without gain or loss along the wire, indicated by $GL\ll1$. Concurrently, we aim to minimize the influence of thermal fluctuations, ensuring that ${\cal N}_{l}\ll 1$. Our goal is to determine the conditions necessary to satisfy these three requirements.

We aim to achieve propagating light without gain or loss, which is possible when $G_u \approx G_l$, leading to $GL \approx 0$. This condition can be satisfied by ensuring that
\begin{equation}\label{Geq0}
\frac{{\cal I}_u}{{\cal I}_l}\approx\frac{1+\Delta_u^2}{1+\Delta_l^2}.
\end{equation}
Additionally, the thermal fluctuation contribution to the signal needs to be significantly less than one. At the waveguide output, at $z=L$, in the limit $GL \ll 1$, and under the condition $\Gamma L/v_g \ll 1$, the thermal contribution is given by
\begin{equation}
{\cal N}_{out}\approx\frac{g^2\Gamma L^2}{v_g^3}\left\{{\cal I}_l\bar{n}+{\cal I}_u(\bar{n}+1)\right\}.
\end{equation}
The contribution of thermal fluctuations to the average number of signal photons at the waveguide output should also be much smaller than one, i.e., ${\cal N}_{out} \ll 1$.

\begin{figure}[t]
\includegraphics[width=0.8\linewidth]{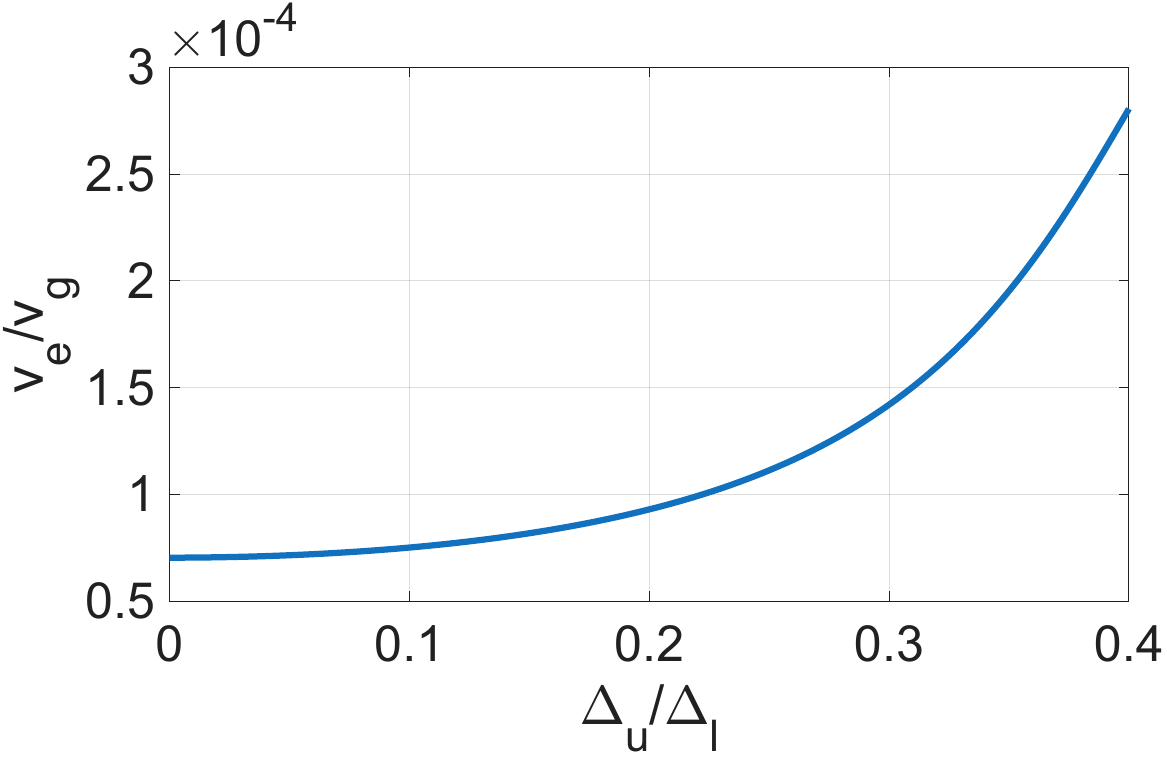}
\caption{The relative effective group velocity $\frac{v_e}{v_g}$ as a function of the relative scaled detuning $\frac{\Delta_u}{\Delta_l}$.}
 \label{Fig10}
\end{figure}

\begin{figure}[t]
\includegraphics[width=0.8\linewidth]{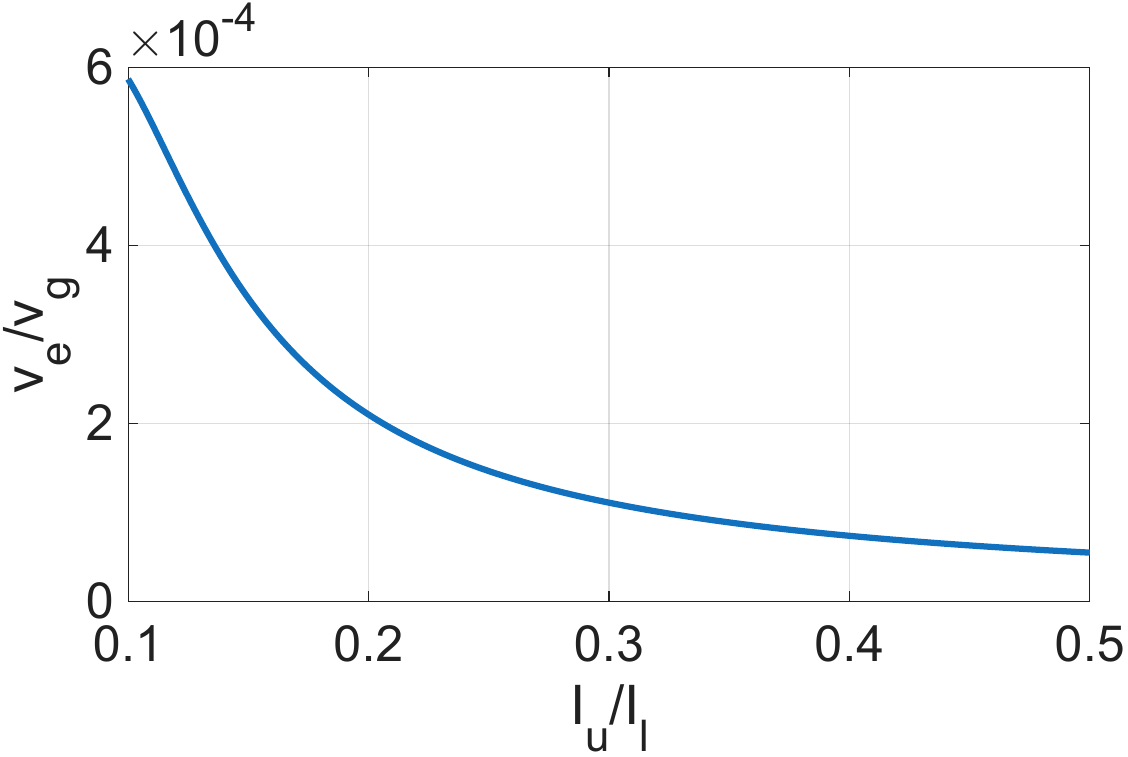}
\caption{The relative effective group velocity $\frac{v_e}{v_g}$ as a function of the relative pump intensity $\frac{{\cal I}_u}{{\cal I}_i}$.}
 \label{Fig11}
\end{figure}

For further analysis of the result, we define the ratios $a = \frac{{\cal I}_u}{{\cal I}_l}$ and $b = \frac{\Delta_u}{\Delta_l}$. We use ${\cal I}_l = {\cal I}$, then ${\cal I}_u = a {\cal I}$, and $\Delta_l = \Delta$ then $\Delta_u = b \Delta$. The requirement (\ref{Geq0}) is written as $\Delta^2 = \frac{1-a}{a-b^2}$. Note that $1 < a < b^2$ or $b^2 < a < 1$. For illustration we use the physical values: $g=10^6$ Hz, 
$\Gamma=10^8$ Hz, $L=1$ cm, and $v_g=10^8$ m/s, with ${\cal I} = 10^8$. We choose $a = b = \frac{1}{4}$ then $\Delta = 2$. We get $\frac{v_e}{v_g} \approx 10^{-4}$, and $\frac{\partial G}{\partial\omega_s} \approx 1.28 \times 10^{-4}$ s/m. For the thermal contribution we get ${\cal N}_{out} \approx 2.8 \times 10^{-3}$. We obtain a slow light with relatively large bandwidth without gain or loss. The signal field operator is given by $\hat{\psi}_{s}(x,t)=\hat{\psi}_{s}^{in}(x-v_gt)e^{-i\kappa x}$, which propagates with a fixed shape and a phase shift $\kappa$ that yields low effective group velocity. Furthermore, the relative effective velocity $\frac{v_e}{v_g}$ from equation (\ref{ve}) is plotted in figure (\ref{Fig10}) as a function of $\Delta_u/\Delta_l$ for $I_u/I_l=1/4$, and in figure (\ref{Fig11}) as a function of $I_u/I_l$ for $\Delta_u/\Delta_l=1/4$.

The time delay is measured by the time of propagation of slow photons along the nanowire. Hence, for $L=10^{-2}$ m and $v_e=10^{4}$ m/s, we have $\Delta t=L/v_e=1~\mu$s. Therefore, one can achieve time delay of the order of the time of storage of the photon-phonon conversion scenario. Moreover one can achieve a longer delay for smaller effective group velocity and longer nanowire. The advantage of the current scenario is in the fact that the time delay is achievable for single photons.

\section{Conclusions}

Realizing the promise of quantum computers depends on developing physical systems to process quantum information. While candidates have been suggested, each with advantages and disadvantages, none fulfill the complete criteria for efficient quantum computing. Photons are naturally non-interacting particles unsuitable for quantum information processing, but they are widely used in quantum communication \cite{Obrien2007}. We introduced nanophotonic structures as strong candidates for the physical implementation of quantum information processing using photons \cite{Zoubi2016,Zoubi2017,Zoubi2020}. Nanostructures made of solid components serve as quantum devices that are easily integrated into on-chip platforms. Photons within nanoscale structures have been shown to strongly interact, making them suitable for quantum information processing, where we reported a significant nonlinear phase shift among photons propagating in nanoscale waveguides by exploiting interactions among photons mediated by vibrational modes and induced through SBS \cite{Zoubi2017}. We used the nonlinear phase among two counter-propagating photons to realize a deterministic quantum logic gate \cite{Zoubi2021}, and implemented photon-phonon interactions to generate a coherent mix of photons and phonons with manifest quantum phenomena \cite{Zoubi2018,Zoubi2019,Zoubi2021,Zoubi2023}. An important step toward interacting photons inside nanowires through SBS is the achievement of relatively slow light that can yield enough time for efficient quantum processing to take
place. We introduced a configuration that slows down photons by several orders of magnitude via SBS involving sound waves in the presence of pump fields \cite{Zoubi2017,Zoubi2024}, in which slow quantum field on the level of single photons can be obtained without gain or loss.

In the present paper we extended the quantum application of nanscale wires as memory devices for the storage of photons. Two scenarios are introduced for the realization of nanowires as storage components of light. The first scenario rest on the the storage of light by converting an optical signal into acoustic excitation inside a waveguide. The technique has been introduced in \cite{Zhu2007}, and realized at nanoscale wire in \cite{Merklein2017}. The process exploits Brillouin scattering where an optical signal is converted to acoustic wave by the interaction with counter propagating strong pump pulse. The signal can be retrieved by sending a second strong pulse after a storage time $\Delta$. Two limitations appear here due to the sound velocity $v_g$ and the acoustic wave damping rate $\Gamma$. The length of the waveguide needs to be enough long for the acoustic wave propagation along the waveguide dring the storage time, that is $L>v_s \Delta$, and the acoustic wave life time is much longer than the storage time, that is $\Delta <1/\Gamma$. The damping rate condition put limitation on the strength of the signal field and forbids one from storing signals at the level of single photons.

The second scenario rest on time delay of the signal field that achieved by slowing down photons. The technique has been introduced by us in \cite{Zoubi2017, Zoubi2024} in exploiting Brillouin scattering of a signal field with two counter propagating pump fields involving acoustic photons. The pump properties serve as control parameters, and slow light with effective group velocity $v_e$ down to the order of the sound velocity can be achieved. The signal can propagates without gain or loss and obeys the condition $L>v_e \Delta$, where the storage time $\Delta$ can be controlled by switching the pump fields on and off. The influence of thermal phonons has been addressed and the condition for their nglegible effect is extracted. The suggested configuration can be adopted for signal fields at the level of single photons, the fact that makes it useful for quantum information processing involving photons.

\section*{Acknowledgment}

The author acknowledges fruitful discussions with Klemens Hammerer from the Institute of Theoretical Physics at Leibniz University Hanover-Germany.


%

\end{document}